\documentclass[fleqn,usenatbib]{mnras}
\usepackage{amssymb}
\usepackage[percent]{overpic}
\usepackage{times}
\usepackage{graphicx}
\usepackage{amsmath}
\usepackage{mathabx}
\usepackage{subfigure}
\usepackage{natbib}
\usepackage{txfonts}
\usepackage[utf8]{inputenc}
\usepackage[overload]{empheq}
\usepackage{lineno}

\title[Kinematic Structure of Radcliffe Wave]
{Discovery of a coherent, wave-like velocity pattern for the Radcliffe Wave} 

\author[G.-X. Li \& B.-Q. Chen]
{Guang-Xing Li$^{1}$\thanks{E-mail:
gxli@ynu.edu.cn, ligx.ngc7293@gmail.com (GXL); bchen@ynu.edu.cn (BQC).}
and Bing-Qiu Chen$^1$\footnotemark[1]
\\
$^{1}$South-Western Institute for Astronomy Research, Yunnan University, Kunming, 650500, P.\,R.\,China\\
}
\begin{document}

\date{Accepted ???. Received ???; in original form ???}

\pagerange{\pageref{firstpage}--\pageref{lastpage}} \pubyear{2022}
\maketitle
\label{firstpage}

\begin{abstract}
  Recently studies discovered that part of the Gould Belt belongs to a 2.7 kpc-long {coherent, thin} wave consisting of
  a chain of clouds, where a  damped undulation pattern has been identified from the spatial arrangement of the clouds.
  We  use the proper motions of  Young Stellar Objects (YSOs) anchored inside the clouds to study the kinematic structure of the Radcliffe Wave in terms of  $v_z$, and identify a damped, wave-like pattern from the $v_z$ space, which we call
  ``velocity undulation".   We propose a new formalism based on the Ensemble Empirical Mode Decomposition (EEMD) to determine the amplitude, period, and phase of the undulation pattern, and find that the spatial and the velocity undulation share an
  almost identical spatial frequency of about 1.5 kpc, and both are damped when measured from one side to the other.  Measured for the first cycle,
  they exhibit a  phase difference of around
  $2\pi/3$. The structure is oscillating around the midplane of
  the Milky Way disk with an amplitude of $\sim\,130\,\pm\,20\,\rm pc$. The vertical extent of the Radcliffe Wave exceeds the thickness of the molecular disk, suggesting that the undulation of the undulation signature might originate from a perturbation,  e.g. the
  passage of a dwarf galaxy.
\end{abstract}

\begin{keywords}
ISM: clouds  -- ISM: structure --  Galaxy: structure -- local interstellar matter  -- stars: kinematics and dynamics
\end{keywords}

\section{Introduction}
Observations have revealed the widespread existence of long ($\sim$ kpc) and
coherent gas filament in the Milky Way disk  {\citep{Li2013, Goodman2014,
2014A&A...568A..73R, Wang2015,2016ApJS..226....9W,2021ApJ...921L..42V}}, which has been reproduced in simulations {
\citep[e.g.][]{2006MNRAS.371.1663D,2015MNRAS.447.3390D,2016MNRAS.455.3640S,2020MNRAS.492.1594S}}.  In recent years, data from the Gaia mission
\citep{2016A&A...595A...1G} have enabled one to map the spatial distributions of
gas in the Solar vicinity with good accuracy. These observations have confirmed
that these long filaments are basic units through which gas in the Milky Way
disk is organized \citep{2019A&A...625A.135L, Chen2019}. Among these objects, a
particularly interesting one is the kpc-scale Radcliffe Wave discovered by
\citet{2020Natur.578..237A}.

Before Gaia, it was widely believed that the ``Gould Belt"
\citep{1874AmJSA...8..325G}, see also \citet[][and references
therein]{2014Ap.....57..583B}  is a ring-like arrangement of molecular gas. It
is only until recently people realized that this structure belongs to a 2.7
kpc-sized, spatially coherent filamentary structure \citep{2020Natur.578..237A}.
In addition to this, \citet{2020Natur.578..237A} have identified an undulation
mode in the $z$ direction with a period of around $ 2 \,\rm  kpc$, the origin
of such an undulating pattern remains undetermined.
  The coherent nature of the object can be further confirmed using radial velocity measurement available for the whole structure \citep{2020Natur.578..237A}.

In this paper, using proper motion data provided by the Gaia mission, we study
the velocity structure of the Young Stellar Objects (YSOs) associated with
clouds in the Radcliffe Wave and report the discovery and subsequent
analyses of a new, {  coherent, wave-like} undulation pattern in the velocity
space.   
\begin{figure*}
  \centering
  \includegraphics[width = 0.8 \textwidth]{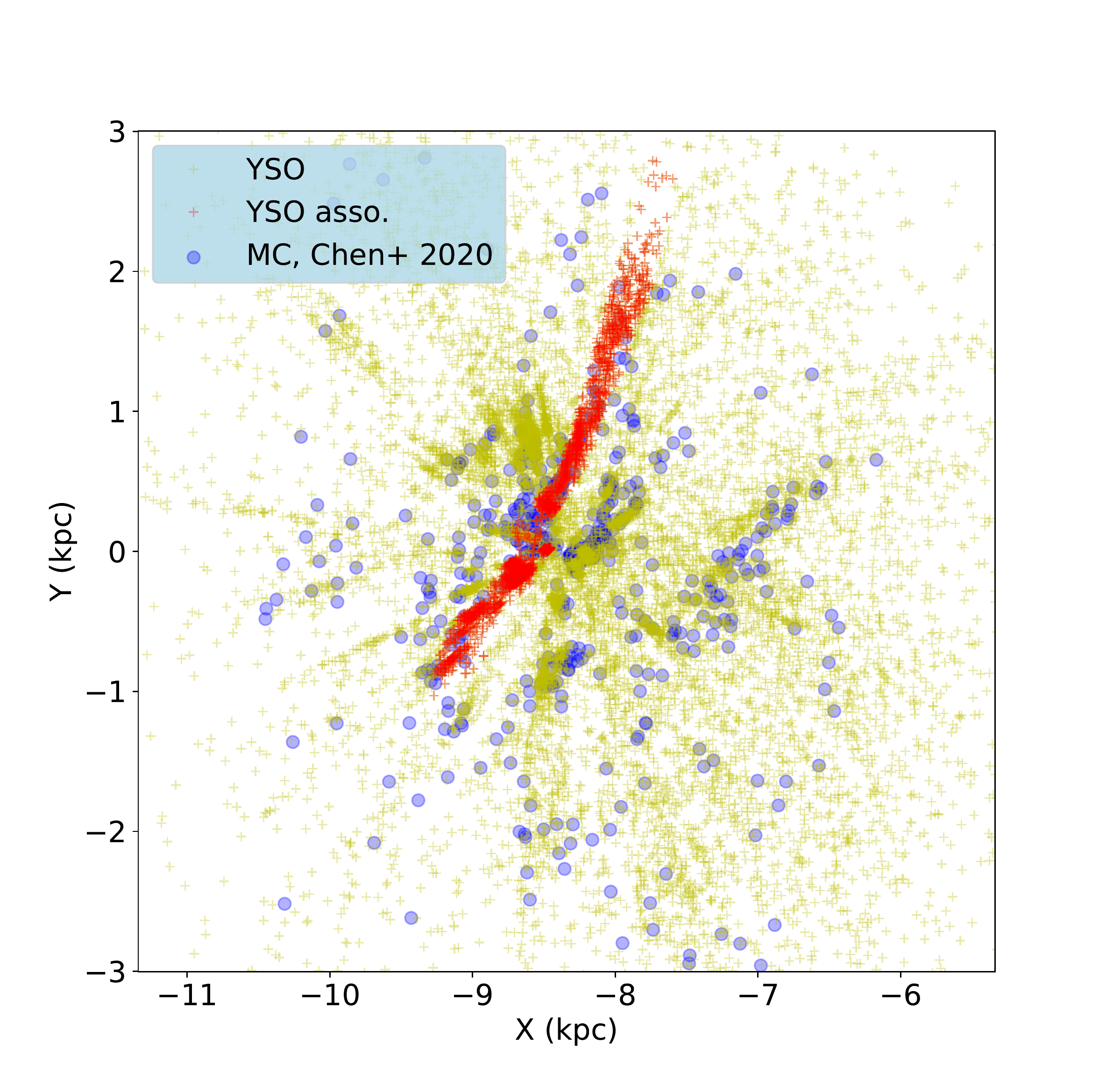}
  \caption { \label{fig:xy} A top-down view of the the Milky Way disk in the Solar vicinity. The Milky Way center is at ($X,~Y$) = $(0,~0)$\,kpc, The Sun sits at $(-8.34, 0)$\,kpc. The blue circles are the catalogue clouds taken from \citet{2020MNRAS.493..351C}. The "+" signs represent the locations of Class~I/II YSOs selected from \citet{2016MNRAS.458.3479M}, where YSOs associated with the Radcliffe Wave are colored in red and the rest in yellow.}
\end{figure*}

\section{Data}

We focus on the kinematic structure measured
along the vertical $z$ direction. The proper motions of
the YSOs that are associated with the Radcliffe Wave are adopted in the current work.  YSOs are stars that were newly
born within clouds and they inherit the motion gas from which they originate.  We use their velocities to study the kinematic structure of the gas they are associated with. We note that although the YSOs inhere the velocity of the gas from which they were born. Because of turbulence, velocity of the gas we observe now can  deviate from the velocity measured at the time when the YSOs are born.  As an estimation, the YSOs we use have an age of $\lesssim 2\rm Myr$ \citep[][and references therein]{2015ApJS..220...11D}.
Assuming that the turbulence in the cloud follows the relation as described by \citet{1981MNRAS.194..809L} where $\sigma_{\rm v}/1 {\rm\,km/ s} \approx (l/1\,\rm pc)^{0.38}$, the crossing time is a function of scale where $t / 1 \,{\rm Myr} = (l / 1 \,{\rm pc})^{0.62}$. Thus, an age difference of 2 Myr translates to velocity dispersion of $1.5\,\rm km\,s^{-1}$, which, to put in simple terms,  is the estimated velocity difference between 
the individual YSOs and the gas they are associated with.

\subsection{YSO selection}

The YSOs are selected from the work of \citet{2016MNRAS.458.3479M}.
Based on the near-infrared photometry from Two Micron
 All-Sky Survey (2MASS; \citealt{Skrutskie2006})  and Wide-field Infrared Survey
 Explorer (WISE; \citealt{Wright2010}) and the 353\,GHz R2.01 Planck dust
 opacity map (\citealt{Planck2014}),  \citet{2016MNRAS.458.3479M} presented  an
 all-sky probabilistic catalog of YSOs using different machine learning
 techniques. We use the sources from the
 \citet{2016MNRAS.458.3479M} catalogue that are classified as  Class I or Class
 II YSOs, which are  the youngest stars that still locate near their birthplaces
 and have similar motions as their parent gases to study the kinematic structure
 of the Radcliffe Wave. Distances and proper motions of these YSOs are obtained by cross-matching with the Gaia Data Release 2 (Gaia DR2; \citealt{Gaia2018}). We further removed sources whose Gaia DR2 parallax
 uncertainties are larger than 20\,per\,cent. This yields 15\,499 Class~I/II
 YSO candidates.

 Fig.~\ref{fig:xy}  plots the spatial distribution of the YSOs in the Galactic disk, where locations of molecular clouds from \citet{2020MNRAS.493..351C} are overlaid.  The selection method can be found in the supplementary material. 

\section{Analysis}
 To study the undulation pattern, we  adopt the same coordinate system as
used by  \citet{2020Natur.578..237A}, where we define a new axis $X'$. The
starting point of this axis anchors at the CMa cloud, which locates at $(X =
-9.3$\,kpc and $Y = -1$\,kpc), and it follows along the filament. Along this axis, in Fig.~\ref{fig:variations}, we plot the spatial variation of $z$ and $v_z$. As a amplitude of the undulation decreases relatively fast  \citep{2020Natur.578..237A}, we focus on its first cycle.

\subsection{Spatial variations of $z$ and $v_z$}
\begin{figure}
  \centering
  \includegraphics[width = 0.49 \textwidth]{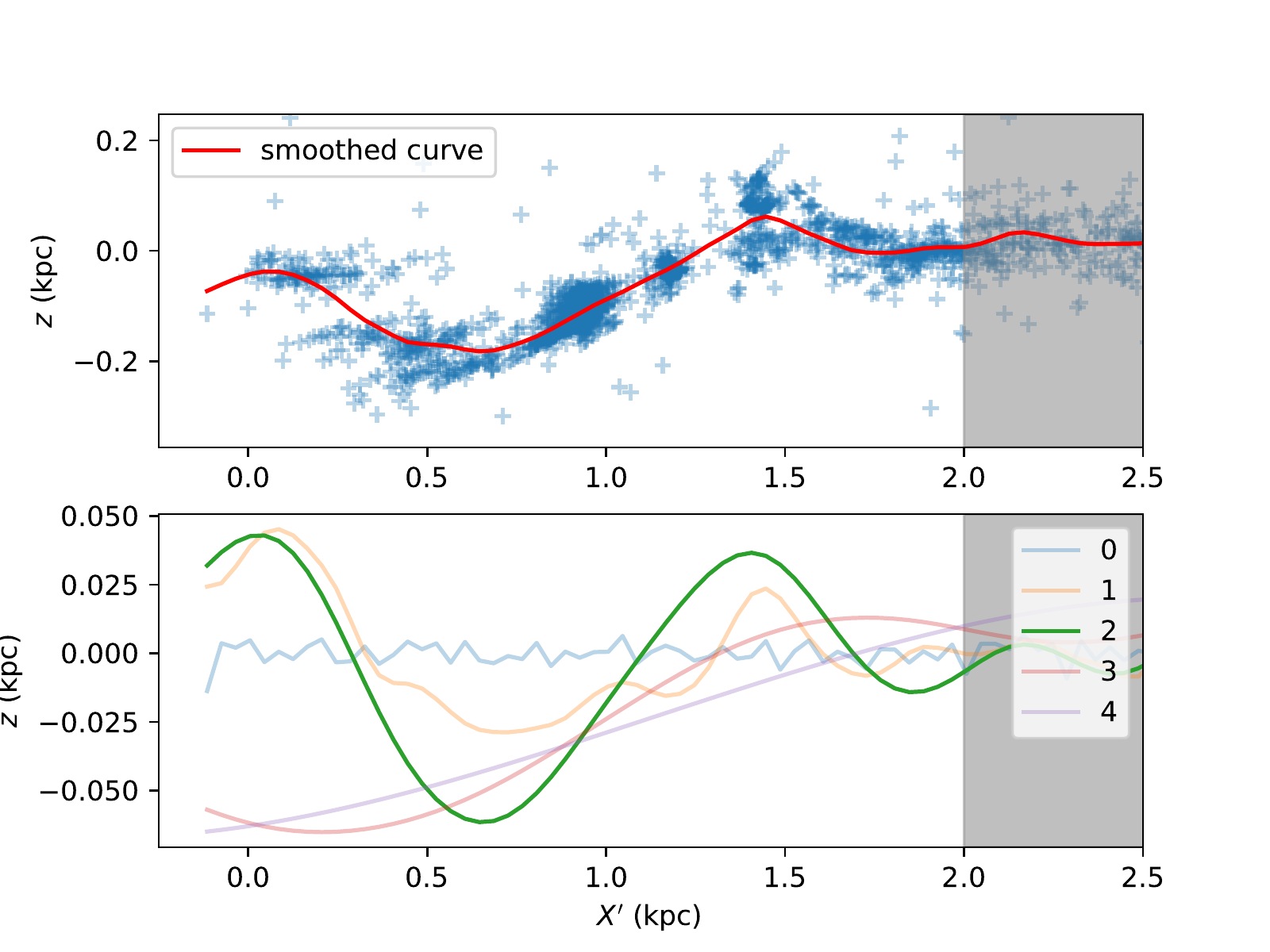}\\
  \includegraphics[width = 0.49 \textwidth]{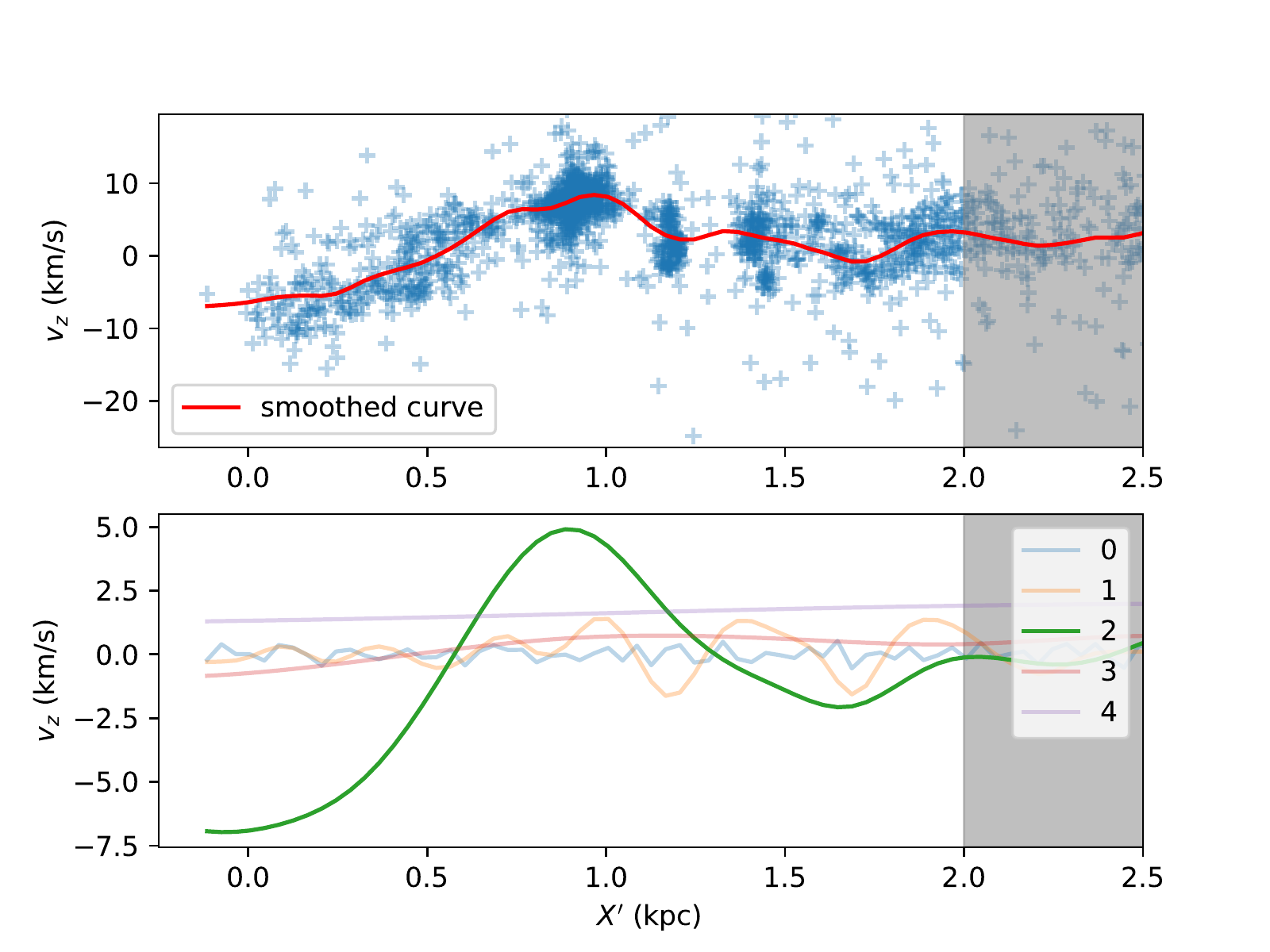}\\
  \caption{Undulation structures of the Radcliffe Wave defined
  along the $X'$ axis. The $X'$ starts from $(X=-7.2$\,kpc and $Y=-1$\,kpc) and follows along
  the filament. We focus on the unshaded region, where $0 < X' < 2 \,\rm kpc$, which corresponds to the first cycle of the undulation. 
   {\it Top panels:} the spatial undulation. The uppermost panel
  plots the spatial displacement from the Galactic disk mid-plane as a function
  of $X'$, where the positions of the YSOs are represented using blue ``+" signs,
  from which we constructed a smoothed profile indicated using the red solid
  line. The second panel plots the result from our  Ensemble Empirical Mode
  Decomposition (EEMD; \citealt{journals/aada/WuH09}) decomposition, where
  different curves represents different Intrinsic Mode Function (IMF) components. The $n=2$ IMF, indicated by
  the green solid line, has the largest amplitude and it corresponds to the
  undulation reported in \citep{2020Natur.578..237A}. {\it Bottom panels:}
  undulation in the velocity space. Third panel: $v_z$ as a function of $X'$, where the 
  velocity of the individual YSOs are presented as blue `+" signs, and the red line is the smoothed curve.
Fourth panel: results from the EEMD decomposition. Curves of different colors represent different IMFs. The  $n=2$ IMF, indicated by
the green solid line, has the largest amplitude and we propose that this is the undulation signal in the velocity space. 
  \label{fig:variations}}
\end{figure}

From Fig,
\ref{fig:variations}, one can identify the sinusoidal modulation in
the $z$ direction as reported by \citet{2020Natur.578..237A}. In addition to this, a
new yet similar feature can also be seen in the $v_z$ plot where the period of the
undulation appears to be similar to the previous one.

\begin{figure}
  \centering
  \includegraphics[width = 0.52 \textwidth]{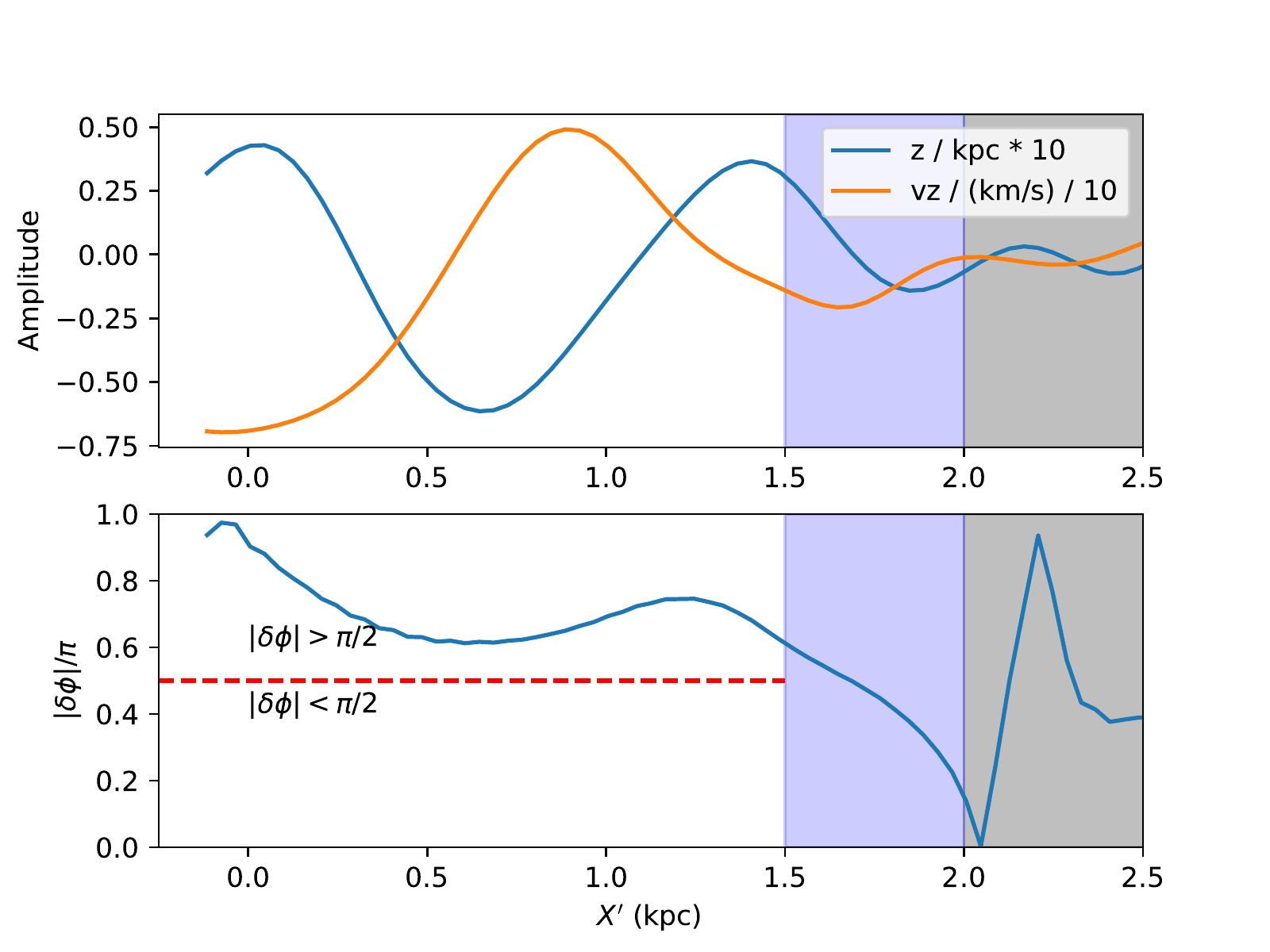}\\
  \caption{\label{fig:phase} Synchronization of the spatial and the velocity undulations. The blue line in the top panel
  represents the $n$=2 IMF of the spatial profile (thus the spatial undulation), and the yellow line in the top panel represents the $n$=2 IMF of the velocity ($v_z$) profile (velocity undulation).
  {\it Bottom panel}: phase difference between the spatial undulation and that of the velocity undulation. They appear to be synchronized within $0{\rm\,kpc}<X'<2 {\rm\,kpc}$. 
 }

\end{figure}

We use the program
\texttt{smoothfit}\footnote{\url{https://pypi.org/project/smoothfit/}.} to
produce smoothed curves which describe the change of $z$ and $v_z$ as a
function of the location $X'$. In this approach, the smoothing is achieved through a
constrained optimization approach. This allows us to average out the measurement  errors of the individual YSOs and focus on the overall structure.   
The smoothed curves are presented in Fig. \ref{fig:variations}.

\subsection{Decomposition of the $z$ and $v_z$ measurements}
We decompose the $z$ and $v_z$ profiles into components of different {  spatial}
frequencies using a \texttt{python} version of Ensemble Empirical Mode
Decomposition (EEMD; \citealt{journals/aada/WuH09}). Being an extension to the 
Empirical Mode Decomposition (EMD; \citealt{1998RSPSA.454..903H}), EEMD is an empirical
method that is designed to decompose a signal into components of different
frequencies, where each component is called an Intrinsic Mode Function (IMF).
These components (IMFs) form a complete and nearly orthogonal basis for the
original signal. Compared to methods such as the wavelet transform, EEMD have a significant advantage in that its performance is robust for nonlinear and non-stationary data. In this paper, we performed EEMD decompositions to both the $z$  and $v_z$ profiles. The results are presented in Fig. \ref{fig:variations}. 

From the EEMD results, 
we identify the $n=2$ IMF from the $z$ profile as the mode that corresponds to the undulation seen in \citet{2020Natur.578..237A}, based on the fact that it is the most prominent IMF, as well as the fact that it has a period of around $\approx 1.5 \,\rm kpc$, which is {  very similar to} the period reported in \citet{2020Natur.578..237A} who  used a damped sinusoidal function to fit the spatial structure. In this paper, we call this feature the ``spatial undulation". 
The fact that we are able to recover this mode from our analysis provides a good validation to our approach. 
Apart from this, the $n= 0, 1$ IMFs represent some low-spatial-frequency components with  lower spatial frequencies, whereas the $n\gtrsim 3$ IMF components represent some higher-order {variations}.


From the decomposition of the $v_z$ profile, we find that the $n=2$ IMF also appears to be the most prominent mode. Similar to the spatial undulation, it also has a period of $\approx 1.5 \,\rm kpc$. In this paper, we called this pattern the ``velocity undulation". From the plots, we measure an amplitude of $75\,\rm pc$ for the spatial undulation and an amplitude of $6\,\rm km\,s^{-1}$ for the velocity undulation. Both the spatial and the velocity undulation appear to be damped, i.e. their amplitudes tend to decrease with increasing $X'$.
The fact that the spatial undulation and velocity undulation has a similar frequency, as well as the fact that both are damped in a similar way strongly imply that they are physically related.

We note that the parameters we derived for the undulation is similar yet not
identical to the values derived by    \citet{2020Natur.578..237A}: we
find an amplitude of $H \approx 130\,\rm pc$ and a period of $p \approx 1.5\rm \,kpc$, whereas
\citet{2020Natur.578..237A} found $H \approx 160$ pc and $p \approx 2 \, \rm  kpc$. The difference is
likely to be caused by the differences in model and method:
The \citet{2020Natur.578..237A} model assumed that the structure is aligned with the disk
mid-plane, whereas in our model we do not impose such a condition and have found that the wave is tilted, e.g.
the $X'=0$ side of the wave is tilted downwards.  {According to our model, a wave crest can be found at $X'=0$, and this is different from \citet{2020Natur.578..237A}. }
The inclusion of the new tiled component
is the main reason why our analyses yield a smaller amplitude and a
smaller period. Nevertheless, we believe that the true period is likely to be
around our value of 1.5\,kpc as this is the common period shared by the spatial
and the velocity undulation. The statistical error and systematics of our analyses
should also be discussed: In the first cycle, our dataset has a YSO density of about
400 per kpc. Therefore, the crest of the wave, which is about 0.3 kpc long, is
sampled by around 120 YSOs, and a typical, 50 pc-sized cloud should be sampled by
$\sim$ 20 YSOs. Assuming a typical distance of $\sim$ 1 kpc, the GAIA distance error is $\sim$ 100 pc per YSO, and we expect a statistical distance error of $\sim$ 22 pc per cloud. Since we are 
only interested in studying the pattern in the $z$ direction, this only amounts to $22\,{\rm pc} / 1\,{\rm kpc} = 2\%$ error.
  Our sample
YSOs have a median parallax error of 0.17 mas/yr, which corresponds to an
velocity error of 0.8 $\rm km\,s^{-1}$ for individual YSOs at 1 kpc,  and a statistical error of
0.2 $\rm km\,s^{-1}$ per cloud. The velocity measurement
 also contain other errors: Due to the lack of radial velocities, what we measured is not strictly $v_z$, but 
rather the velocity projected onto an equi-distance sphere from the earth. 
This velocity error caused by projection can be as large as 20\,per\,cent (Sec.~\ref{sec:vz:comp}).

Other factors such as intrinsic thickness and intrinsic velocity dispersion can also be important. In our formalism, these are contained in the n= 1, 2 IMFs of the EEMD results. This is true both for the decomposition of the spatial undulation signal and the velocity one. From these, we estimate an error of about 6 pc for the spatial undulation signal and an error of around 1 $\rm km s^{-1}$ for the velocity undulation signal, for the first cycle.


\subsection{Determining the phase difference}
We use Hilbert transform \citep[e.g.][]{Hilbert} to study phase difference between the spatial and the velocity undulation. The Hilbert transform is a linear transform typically applied to real-valued signals to derive the real-time evolution of the phase and the amplitude. 

The results are present in Fig. \ref{fig:phase}. Within $0\,{\rm kpc}<X'<1.5\,{\rm kpc}$, the spatial and velocity undulation have a phase difference of $\approx  2/3 \pi$. They start to de-synchronize from $X' = 1.5\,{\rm kpc}$. However, this might be due to the fact that the error of proper motion has increased and the fact that the filament is sparely sampled by YSOs for $X' > 1.5$\,kpc.


\section{Discussion}
\subsection{Phase shift}
The most striking discovery from our research is the coherent, wave-like undulation in the velocity space. The spatial frequency of the velocity undulation is similar to that of the spatial undulation, and these undulations exhibit a phase difference of $\approx 2/3 \pi$. This value places our target between being traveling-wave-like 
(of phase difference $\pm\,\pi/2$) and being  standing-wave-like ($\pm \pi$).  This synchronized pattern provides insights into  the nature of the structure.


\subsection{Overall Amplitude}
We estimate the full extent within which the oscillation occurs. We assume that the oscillation has a total height of $H_{\rm tot}$, and the velocity of the filament comes from the kinetic energy released when it falls from $H = H_{\rm tot}$ to its current height $H_0$. 

We  use the following equation to estimate th full amplitude (see online supplementary material for more details):
\begin{equation}
  H_{\rm tot} \approx \Big{(} \frac{v_0^2}{4 \pi G \rho_*} + H_0^2  \Big{)}^{1/2}  \approx 130 \pm 20 \,\rm pc ,
\end{equation}
 where $\rho_*$ is the mass density from stars and dark matter, $v_0$ is the amplitude of the velocity undulation and $H_0$ is the amplitude of the spatial undulation. 
  We use $v_{\rm 0} = 6\pm 1\,\rm km\,s^{-1}$, $H_0 = 70\pm 6\,\rm pc$ and
$\rho_* = 0.05\,\rm M_{\odot}\,\rm pc^{-3}$ \citep{2003A&A...409..523R}. The oscillation should have an amplitude
of $130\,\rm pc$, meaning that the whole structure is probably oscillating
within a vertical range of around $260\,\rm pc$. We can also estimate the maximum velocity the gas would reach when the it passes through the Galactic disk, which is
\begin{equation}
  v_{\rm max} = \sqrt{4 \pi G \rho_* H_0^2} \approx 7\pm 0.6 \,\rm km\,s^{-1} \,.
\end{equation}


The Radcliffe Wave has an estimated height of 130\,pc (or a spatial
extent of 260\,pc). Thus oscillation is not confined to the thick molecular disk
(FWHM = 120\.pc; \citealt{2015ARA&A..53..583H}). 

\subsection{Spatial inhomogeneity and Possible Origin}
In line with previous studies, our results suggest that the structure of the
 Radcliffe Wave is spatially inhomogeneous: not only do the
undulation appears to be damped such that the amplitude decreases with
increasing $X'$, the Radcliffe
Wave is also tilted, e.g. the spatial offset measured with respect to the $b=0$ plane
decreases by around 100 pc as $X'$ increases from 0 to 1.5 kpc. Should the Gould
Belt Radcliffe Wave be created by a perturbation, the source should be near $X' =0$.  
   
We combine these undulation
 patterns to determine the nature of the undulation pattern. If the undulation signature evolves in  a way that is similar to that of the travelling wave, we can determine
 the direction of the travel. At $X'=$ 1 kpc, seen from
 Fig. \ref{fig:phase}, the wave has reached maximum velocity, and are moving
 upwards. The corresponding spatial pattern then implies that the wave would be
 propagating leftwards. This scenario is unlikely as the wave is damped when
 measured from the left to the right. Should the undulation pattern be caused by
 a perturbation occurring at $X'=0$, we expect the wave to travel towards the
 right.  
Because of this,  the standing wave {might be} a slightly favoured
scenario. We note, however that as the phase shift changes as a function of the
location, the exact nature of the undulation remains ambiguous.

Recent data from the GAIA
satellite have revealed many phase-space patterns, which can be explained by passages of dwarf galaxies \citep{2018Natur.561..360A,2019MNRAS.489.4962K}. These dwarf galaxies are abundant \citep{2019ARA&A..57..375S},  thus these passages should occur regularly. The undulation may be caused by one of these passages.


\section{Conclusion}
We perform a study of the kinematic structure of the newly-discovered  Radcliffe Wave -- a 2.7-kpc long wave-like arrangement of molecular gas.  We propose to use the velocities of YSOs, which are young accreting stars that were newly born within molecular clouds, to trace their kinematic structure.
Using our new data,
 we identify a similar pattern from the velocity along the Galactic $z$ direction, which might share a common origin as the undulation signature. We call our new coherent, wave-like velocity pattern the ``velocity undulation". 

\citet{2020Natur.578..237A} identified an undulation of a period of $\approx 2\,\rm kpc$ and an amplitude of 160 pc from the spatial arrangement of gas. Because of the differences in the models,  our analyses yield slightly different results:  The left side of the wave is tilted upwards. The wave has a period of 1.5 kpc, a height of $H = 70\pm 6 \,\rm pc$, and a velocity of $7\pm 1\,\rm km/s$.
 The velocity undulation is damped similarly to that of the spatial undulation. The spatial and the velocity undulation share an almost identical period of $\approx 1.5\,\rm kpc$,  and they exhibit a phase difference of $\approx 2/3 \,\pi$, which places our objects in the middle of being standing-wave-like and traveling-wave-like. We note at $X' = 1 \,\rm kpc$, the gas is sitting close to the disk midplane and is moving upwards (with a positive $v_z$), consistent with the structure behaving either like a standing wave, or a wave traveling towards the left.

From a joint analysis of the spatial and velocity undulation, we estimate that
 the wave is possibly oscillating around the Galactic disk mid-plane with an amplitude of $\approx 130\,\pm 20 \,\rm pc$ (e.g. the gas should oscillate within $z = \pm 130\,\rm pc$). The scale on which the oscillation occurs is larger than the mean thickness of the thin molecular disk in the solar vicinity ($H = 73\,\rm pc$), indicating that the wave is moderately perturbed.  This, combined with the spatial inhomogeneity suggests that the undulation seen in the  Radcliffe Wave can be caused by some perturbations e.g. the passage of a dwarf galaxy.

\section*{Acknowledgements}
This work is partially supported by National Natural Science Foundation of China grants No.\,W820301904, 12173034 and 11833006,
and Yunnan University grant No.~C176220100007 and 	No.~C176220100028. We acknowledge the science research
grants from the China Manned Space Project with NO.\,CMS-CSST-2021-A09, CMS-CSST-2021-A08 and CMS-CSST-2021-B03. 

This work presents results from the European Space Agency (ESA) space mission Gaia. Gaia data are being processed by the Gaia Data Processing and Analysis Consortium (DPAC). Funding for the DPAC is provided by national institutions, in particular the institutions participating in the Gaia MultiLateral Agreement (MLA). The Gaia mission website is https://www.cosmos.esa.int/gaia. The Gaia archive website is https://archives.esac.esa.int/gaia.

 
\section*{Data availability}

The paper makes use of publicly-available data presented in \citet{2016MNRAS.458.3479M,Gaia2018}.

\bibliographystyle{mnras}
\bibliography{paper}

\begin{thebibliography}{}
\makeatletter
\relax
\def\mn@urlcharsother{\let\do\@makeother \do\$\do\&\do\#\do\^\do\_\do\%\do\~}
\def\mn@doi{\begingroup\mn@urlcharsother \@ifnextchar [ {\mn@doi@}
  {\mn@doi@[]}}
\def\mn@doi@[#1]#2{\def\@tempa{#1}\ifx\@tempa\@empty \href
  {http://dx.doi.org/#2} {doi:#2}\else \href {http://dx.doi.org/#2} {#1}\fi
  \endgroup}
\def\mn@eprint#1#2{\mn@eprint@#1:#2::\@nil}
\def\mn@eprint@arXiv#1{\href {http://arxiv.org/abs/#1} {{\tt arXiv:#1}}}
\def\mn@eprint@dblp#1{\href {http://dblp.uni-trier.de/rec/bibtex/#1.xml}
  {dblp:#1}}
\def\mn@eprint@#1:#2:#3:#4\@nil{\def\@tempa {#1}\def\@tempb {#2}\def\@tempc
  {#3}\ifx \@tempc \@empty \let \@tempc \@tempb \let \@tempb \@tempa \fi \ifx
  \@tempb \@empty \def\@tempb {arXiv}\fi \@ifundefined
  {mn@eprint@\@tempb}{\@tempb:\@tempc}{\expandafter \expandafter \csname
  mn@eprint@\@tempb\endcsname \expandafter{\@tempc}}}

\bibitem[\protect\citeauthoryear{{Alves} et~al.,}{{Alves}
  et~al.}{2020}]{2020Natur.578..237A}
{Alves} J.,  et~al., 2020, \mn@doi [\nat] {10.1038/s41586-019-1874-z}, \href
  {https://ui.adsabs.harvard.edu/abs/2020Natur.578..237A} {578, 237}

\bibitem[\protect\citeauthoryear{{Antoja} et~al.,}{{Antoja}
  et~al.}{2018}]{2018Natur.561..360A}
{Antoja} T.,  et~al., 2018, \mn@doi [\nat] {10.1038/s41586-018-0510-7}, \href
  {https://ui.adsabs.harvard.edu/abs/2018Natur.561..360A} {561, 360}

\bibitem[\protect\citeauthoryear{{Beaumont}, {Goodman}  \&
  {Greenfield}}{{Beaumont} et~al.}{2015}]{2015ASPC..495..101B}
{Beaumont} C.,  {Goodman} A.,   {Greenfield} P.,  2015, in {Taylor} A.~R.,
  {Rosolowsky} E.,  eds,  Astronomical Society of the Pacific Conference Series
  Vol. 495, Astronomical Data Analysis Software an Systems XXIV (ADASS XXIV).
  p.~101

\bibitem[\protect\citeauthoryear{{Bobylev}}{{Bobylev}}{2014}]{2014Ap.....57..583B}
{Bobylev} V.~V.,  2014, \mn@doi [Astrophysics] {10.1007/s10511-014-9360-7},
  \href {https://ui.adsabs.harvard.edu/abs/2014Ap.....57..583B} {57, 583}

\bibitem[\protect\citeauthoryear{{Bovy}}{{Bovy}}{2015}]{2015ApJS..216...29B}
{Bovy} J.,  2015, \mn@doi [\apjs] {10.1088/0067-0049/216/2/29}, \href
  {https://ui.adsabs.harvard.edu/abs/2015ApJS..216...29B} {216, 29}

\bibitem[\protect\citeauthoryear{{Chen} et~al.,}{{Chen}
  et~al.}{2019}]{Chen2019}
{Chen} B.-Q.,  et~al., 2019, \mn@doi [\mnras] {10.1093/mnras/sty3341}, \href
  {http://adsabs.harvard.edu/abs/2019MNRAS.483.4277C} {483, 4277}

\bibitem[\protect\citeauthoryear{{Chen} et~al.,}{{Chen}
  et~al.}{2020}]{2020MNRAS.493..351C}
{Chen} B.~Q.,  et~al., 2020, \mn@doi [\mnras] {10.1093/mnras/staa235}, \href
  {https://ui.adsabs.harvard.edu/abs/2020MNRAS.493..351C} {493, 351}

\bibitem[\protect\citeauthoryear{{Dobbs}}{{Dobbs}}{2015}]{2015MNRAS.447.3390D}
{Dobbs} C.~L.,  2015, \mn@doi [\mnras] {10.1093/mnras/stu2585}, \href
  {https://ui.adsabs.harvard.edu/abs/2015MNRAS.447.3390D} {447, 3390}

\bibitem[\protect\citeauthoryear{{Dobbs}, {Bonnell}  \& {Pringle}}{{Dobbs}
  et~al.}{2006}]{2006MNRAS.371.1663D}
{Dobbs} C.~L.,  {Bonnell} I.~A.,   {Pringle} J.~E.,  2006, \mn@doi [\mnras]
  {10.1111/j.1365-2966.2006.10794.x}, \href
  {http://adsabs.harvard.edu/abs/2006MNRAS.371.1663D} {371, 1663}

\bibitem[\protect\citeauthoryear{{Dunham} et~al.,}{{Dunham}
  et~al.}{2015}]{2015ApJS..220...11D}
{Dunham} M.~M.,  et~al., 2015, \mn@doi [\apjs] {10.1088/0067-0049/220/1/11},
  \href {https://ui.adsabs.harvard.edu/abs/2015ApJS..220...11D} {220, 11}

\bibitem[\protect\citeauthoryear{{Gaia Collaboration} et~al.,}{{Gaia
  Collaboration} et~al.}{2016}]{2016A&A...595A...1G}
{Gaia Collaboration} et~al., 2016, \mn@doi [\aap]
  {10.1051/0004-6361/201629272}, \href
  {https://ui.adsabs.harvard.edu/abs/2016A&A...595A...1G} {595, A1}

\bibitem[\protect\citeauthoryear{{Gaia Collaboration} et~al.,}{{Gaia
  Collaboration} et~al.}{2018}]{Gaia2018}
{Gaia Collaboration} et~al., 2018, \mn@doi [\aap]
  {10.1051/0004-6361/201833051}, \href
  {http://adsabs.harvard.edu/abs/2018A%26A...616A...1G} {616, A1}

\bibitem[\protect\citeauthoryear{{Goodman} et~al.,}{{Goodman}
  et~al.}{2014}]{Goodman2014}
{Goodman} A.~A.,  et~al., 2014, \mn@doi [\apj] {10.1088/0004-637X/797/1/53},
  \href {http://adsabs.harvard.edu/abs/2014ApJ...797...53G} {797, 53}

\bibitem[\protect\citeauthoryear{{Gould}}{{Gould}}{1874}]{1874AmJSA...8..325G}
{Gould} B.~A.,  1874, American Journal of Science and Arts, \href
  {https://ui.adsabs.harvard.edu/abs/1874AmJSA...8..325G} {8, 325}

\bibitem[\protect\citeauthoryear{Hahn}{Hahn}{1996}]{Hilbert}
Hahn S.~L.,  1996, Hilbert Transforms in Signal Processing.
Artech House Publishers

\bibitem[\protect\citeauthoryear{{Heyer} \& {Dame}}{{Heyer} \&
  {Dame}}{2015}]{2015ARA&A..53..583H}
{Heyer} M.,  {Dame} T.~M.,  2015, \mn@doi [\araa]
  {10.1146/annurev-astro-082214-122324}, \href
  {https://ui.adsabs.harvard.edu/abs/2015ARA&A..53..583H} {53, 583}

\bibitem[\protect\citeauthoryear{{Huang} et~al.,}{{Huang}
  et~al.}{1998}]{1998RSPSA.454..903H}
{Huang} N.~E.,  et~al., 1998, \mn@doi [Proceedings of the Royal Society of
  London Series A] {10.1098/rspa.1998.0193}, \href
  {https://ui.adsabs.harvard.edu/abs/1998RSPSA.454..903H} {454, 903}

\bibitem[\protect\citeauthoryear{{Khanna} et~al.,}{{Khanna}
  et~al.}{2019}]{2019MNRAS.489.4962K}
{Khanna} S.,  et~al., 2019, \mn@doi [\mnras] {10.1093/mnras/stz2462}, \href
  {https://ui.adsabs.harvard.edu/abs/2019MNRAS.489.4962K} {489, 4962}

\bibitem[\protect\citeauthoryear{{Lallement}, {Babusiaux}, {Vergely}, {Katz},
  {Arenou}, {Valette}, {Hottier}  \& {Capitanio}}{{Lallement}
  et~al.}{2019}]{2019A&A...625A.135L}
{Lallement} R.,  {Babusiaux} C.,  {Vergely} J.~L.,  {Katz} D.,  {Arenou} F.,
  {Valette} B.,  {Hottier} C.,   {Capitanio} L.,  2019, \mn@doi [\aap]
  {10.1051/0004-6361/201834695}, \href
  {https://ui.adsabs.harvard.edu/abs/2019A&A...625A.135L} {625, A135}

\bibitem[\protect\citeauthoryear{{Larson}}{{Larson}}{1981}]{1981MNRAS.194..809L}
{Larson} R.~B.,  1981, \mn@doi [\mnras] {10.1093/mnras/194.4.809}, \href
  {https://ui.adsabs.harvard.edu/abs/1981MNRAS.194..809L} {194, 809}

\bibitem[\protect\citeauthoryear{{Li}, {Wyrowski}, {Menten}  \&
  {Belloche}}{{Li} et~al.}{2013}]{Li2013}
{Li} G.-X.,  {Wyrowski} F.,  {Menten} K.,   {Belloche} A.,  2013, \mn@doi
  [\aap] {10.1051/0004-6361/201322411}, \href
  {http://adsabs.harvard.edu/abs/2013A%26A...559A..34L} {559, A34}

\bibitem[\protect\citeauthoryear{Luo et~al.,}{Luo et~al.}{2015}]{Luo_2015}
Luo A.-L.,  et~al., 2015, \mn@doi [Research in Astronomy and Astrophysics]
  {10.1088/1674-4527/15/8/002}, 15, 1095

\bibitem[\protect\citeauthoryear{{Marton}, {T{\'o}th}, {Paladini}, {Kun},
  {Zahorecz}, {McGehee}  \& {Kiss}}{{Marton}
  et~al.}{2016}]{2016MNRAS.458.3479M}
{Marton} G.,  {T{\'o}th} L.~V.,  {Paladini} R.,  {Kun} M.,  {Zahorecz} S.,
  {McGehee} P.,   {Kiss} C.,  2016, \mn@doi [\mnras] {10.1093/mnras/stw398},
  \href {https://ui.adsabs.harvard.edu/abs/2016MNRAS.458.3479M} {458, 3479}

\bibitem[\protect\citeauthoryear{{Planck Collaboration} et~al.,}{{Planck
  Collaboration} et~al.}{2014}]{Planck2014}
{Planck Collaboration} et~al., 2014, \mn@doi [\aap]
  {10.1051/0004-6361/201323195}, \href
  {http://adsabs.harvard.edu/abs/2014A%26A...571A..11P} {571, A11}

\bibitem[\protect\citeauthoryear{{Planck Collaboration} et~al.,}{{Planck
  Collaboration} et~al.}{2020}]{2020A&A...641A...1P}
{Planck Collaboration} et~al., 2020, \mn@doi [\aap]
  {10.1051/0004-6361/201833880}, \href
  {https://ui.adsabs.harvard.edu/abs/2020A&A...641A...1P} {641, A1}

\bibitem[\protect\citeauthoryear{{Ragan}, {Henning}, {Tackenberg}, {Beuther},
  {Johnston}, {Kainulainen}  \& {Linz}}{{Ragan}
  et~al.}{2014}]{2014A&A...568A..73R}
{Ragan} S.~E.,  {Henning} T.,  {Tackenberg} J.,  {Beuther} H.,  {Johnston}
  K.~G.,  {Kainulainen} J.,   {Linz} H.,  2014, \mn@doi [\aap]
  {10.1051/0004-6361/201423401}, \href
  {http://adsabs.harvard.edu/abs/2014A%26A...568A..73R} {568, A73}

\bibitem[\protect\citeauthoryear{{Robin}, {Reyl{\'e}}, {Derri{\`e}re}  \&
  {Picaud}}{{Robin} et~al.}{2003}]{2003A&A...409..523R}
{Robin} A.~C.,  {Reyl{\'e}} C.,  {Derri{\`e}re} S.,   {Picaud} S.,  2003,
  \mn@doi [\aap] {10.1051/0004-6361:20031117}, \href
  {https://ui.adsabs.harvard.edu/abs/2003A&A...409..523R} {409, 523}

\bibitem[\protect\citeauthoryear{Robitaille, Beaumont, Qian, Borkin  \&
  Goodman}{Robitaille et~al.}{2017}]{robitaille_thomas_2017_1237692}
Robitaille T.,  Beaumont C.,  Qian P.,  Borkin M.,   Goodman A.,  2017, glueviz
  v0.13.1: multidimensional data exploration, \mn@doi{10.5281/zenodo.1237692},
  \url {https://doi.org/10.5281/zenodo.1237692}

\bibitem[\protect\citeauthoryear{{Simon}}{{Simon}}{2019}]{2019ARA&A..57..375S}
{Simon} J.~D.,  2019, \mn@doi [\araa] {10.1146/annurev-astro-091918-104453},
  \href {https://ui.adsabs.harvard.edu/abs/2019ARA&A..57..375S} {57, 375}

\bibitem[\protect\citeauthoryear{{Skrutskie} et~al.,}{{Skrutskie}
  et~al.}{2006}]{Skrutskie2006}
{Skrutskie} M.~F.,  et~al., 2006, \mn@doi [\aj] {10.1086/498708}, \href
  {https://ui.adsabs.harvard.edu/abs/2006AJ....131.1163S} {131, 1163}

\bibitem[\protect\citeauthoryear{{Smith}, {Glover}, {Klessen}  \&
  {Fuller}}{{Smith} et~al.}{2016}]{2016MNRAS.455.3640S}
{Smith} R.~J.,  {Glover} S. C.~O.,  {Klessen} R.~S.,   {Fuller} G.~A.,  2016,
  \mn@doi [\mnras] {10.1093/mnras/stv2559}, \href
  {https://ui.adsabs.harvard.edu/abs/2016MNRAS.455.3640S} {455, 3640}

\bibitem[\protect\citeauthoryear{{Smith} et~al.,}{{Smith}
  et~al.}{2020}]{2020MNRAS.492.1594S}
{Smith} R.~J.,  et~al., 2020, \mn@doi [\mnras] {10.1093/mnras/stz3328}, \href
  {https://ui.adsabs.harvard.edu/abs/2020MNRAS.492.1594S} {492, 1594}

\bibitem[\protect\citeauthoryear{{Veena} et~al.,}{{Veena}
  et~al.}{2021}]{2021ApJ...921L..42V}
{Veena} V.~S.,  et~al., 2021, \mn@doi [\apjl] {10.3847/2041-8213/ac341f}, \href
  {https://ui.adsabs.harvard.edu/abs/2021ApJ...921L..42V} {921, L42}

\bibitem[\protect\citeauthoryear{{Wang}, {Testi}, {Ginsburg}, {Walmsley},
  {Molinari}  \& {Schisano}}{{Wang} et~al.}{2015}]{Wang2015}
{Wang} K.,  {Testi} L.,  {Ginsburg} A.,  {Walmsley} C.~M.,  {Molinari} S.,
  {Schisano} E.,  2015, \mn@doi [\mnras] {10.1093/mnras/stv735}, \href
  {http://adsabs.harvard.edu/abs/2015MNRAS.450.4043W} {450, 4043}

\bibitem[\protect\citeauthoryear{{Wang}, {Testi}, {Burkert}, {Walmsley},
  {Beuther}  \& {Henning}}{{Wang} et~al.}{2016}]{2016ApJS..226....9W}
{Wang} K.,  {Testi} L.,  {Burkert} A.,  {Walmsley} C.~M.,  {Beuther} H.,
  {Henning} T.,  2016, \mn@doi [\apjs] {10.3847/0067-0049/226/1/9}, \href
  {https://ui.adsabs.harvard.edu/abs/2016ApJS..226....9W} {226, 9}

\bibitem[\protect\citeauthoryear{{Wright} et~al.,}{{Wright}
  et~al.}{2010}]{Wright2010}
{Wright} E.~L.,  et~al., 2010, \mn@doi [\aj] {10.1088/0004-6256/140/6/1868},
  \href {https://ui.adsabs.harvard.edu/abs/2010AJ....140.1868W} {140, 1868}

\bibitem[\protect\citeauthoryear{Wu \& Huang}{Wu \&
  Huang}{2009}]{journals/aada/WuH09}
Wu Z.,  Huang N.~E.,  2009, Adv. Data Sci. Adapt. Anal., 1, 1

\makeatother
\end{thebibliography}

\clearpage
\appendix
 \section*{Estimation of total amplitude (scale-height)} \label{sec:total}
The total height within which the oscillation occurs is determined by its total energy, which is the sum of  the kinetic energy and the potential energy. {To better understand the oscillation,}
 we assume that the Galactic disk can be approximated as a slab of a constant density $\rho_*$. Starting from the mid-plane of the slab, using the Poisson equation $\nabla^2 \phi = 4 \pi G \rho$, the acceleration can be computed as
\begin{equation}
  \frac{{\rm d} a_z}{{\rm d} z} = - 4 \pi G \rho ,
\end{equation}
thus $a_z = - 4 \pi G \rho z$  and $e_{\rm g} = 2 \pi G \rho z^2$ where $e_{\rm g}$ is the gravitational energy per unit mass.  {We assume $\rho$ to be a constant, for simplicity.} The total kinetic energy of the oscillation is 
\begin{equation}
  e_{\rm tot, z} =  2 \pi G \rho_* z^2 + \frac{1}{2} v_z^2,
  \end{equation} 
from which we can infer the total height
\begin{equation}
  H_{\rm tot} = \big{(} \frac{e_{\rm tot}}{2 \pi G \rho_*}  \big{)}^{1/2}  =   \big{(}  z^2 + \frac{v_z^2}{4 \pi G \rho_*} \big{)}^{1/2} ,
 \end{equation}
where $z$ is the current height and $v_z$ is the current vertical velocity. The maximum velocity (velocity the gas  should contain when it has fallen to $z= 0$) should be
\begin{equation}
  v_{\rm max} = (2\,e_{\rm tot})^{1/2} = (4 \pi G \rho_* z^2 + v_z^2 )^{1/2} = (4 \pi G \rho_* )^{1/2}  H_{\rm max}\,.
\end{equation}

\section*{Determination of YSO memberships }
\begin{figure*}
\includegraphics[width = 0.9 \textwidth]{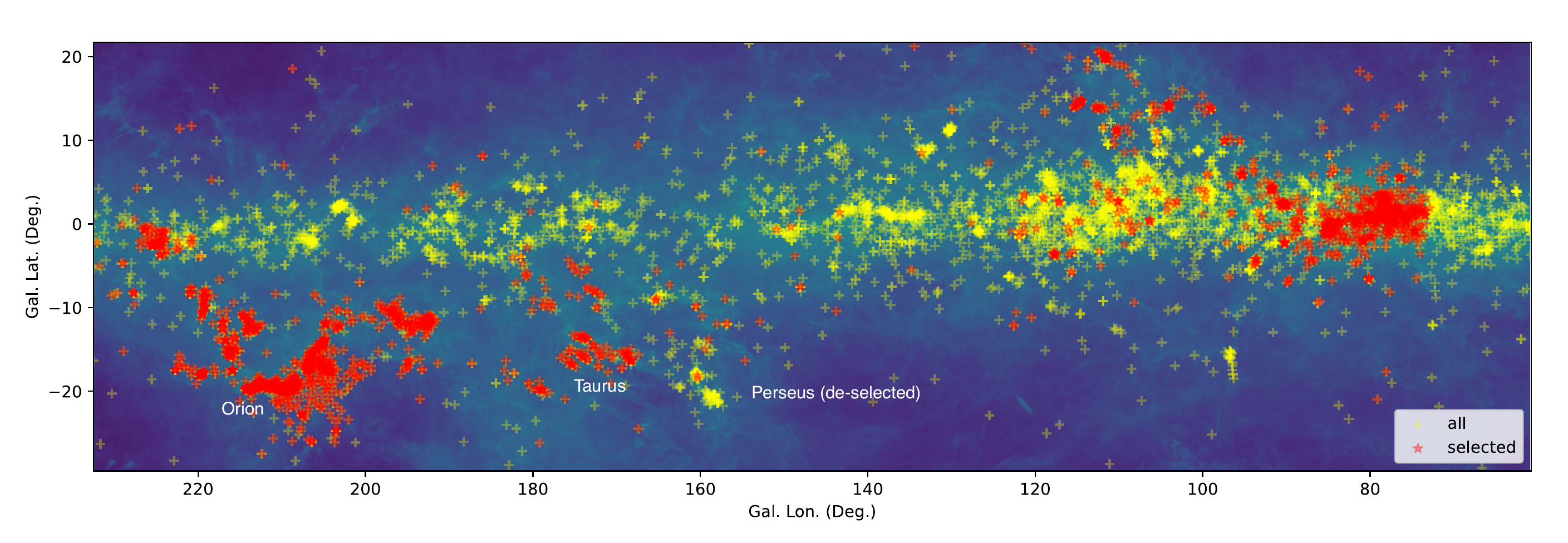}
\caption{\label{fig:selection}. Locations of YSOs on the sky. The locations of our sample YSOs are plotted in yellow, and those associated with the Radcliffe Wave are colored in red. The backgrounnd is a map of 870$\mu m$ continuum obtained by the Planck Satellite \citep{2020A&A...641A...1P}, where the distribution of gas can be traced from the emission by dust. Locations of a few famous molecular clouds are indicated. }
 \end{figure*}

The YSOs associated with the Radcliffe Wave are selected in the Galactic $X$-$Y$ plane manually  with the help of the \texttt{Glue} software \citep{2015ASPC..495..101B,robitaille_thomas_2017_1237692}. {Since the spatial distribution of the YSOs is similar to that of the gas, we select YSOs associated with the Radcliffe wave by drawing a polygon at the Galactic $X$-$Y$ plane. The results are varied by visual inspections afterward. }
 In addition to this, we have de-selected  YSOs associated with the Perseus molecular clouds from the Galactic $l$-$b$ plane since as they exhibit significant spreads, which can be seen in Fig.~2 of \citet{2020Natur.578..237A}. {A plot of all the number YSOs as well as those associated with the Radcliffe Wave can be found in Fig. \ref{fig:selection}.}

\section*{Computation of $v_z$} \label{sec:vz:comp}
\begin{figure}
  \centering
  \includegraphics[width = 0.5 \textwidth]{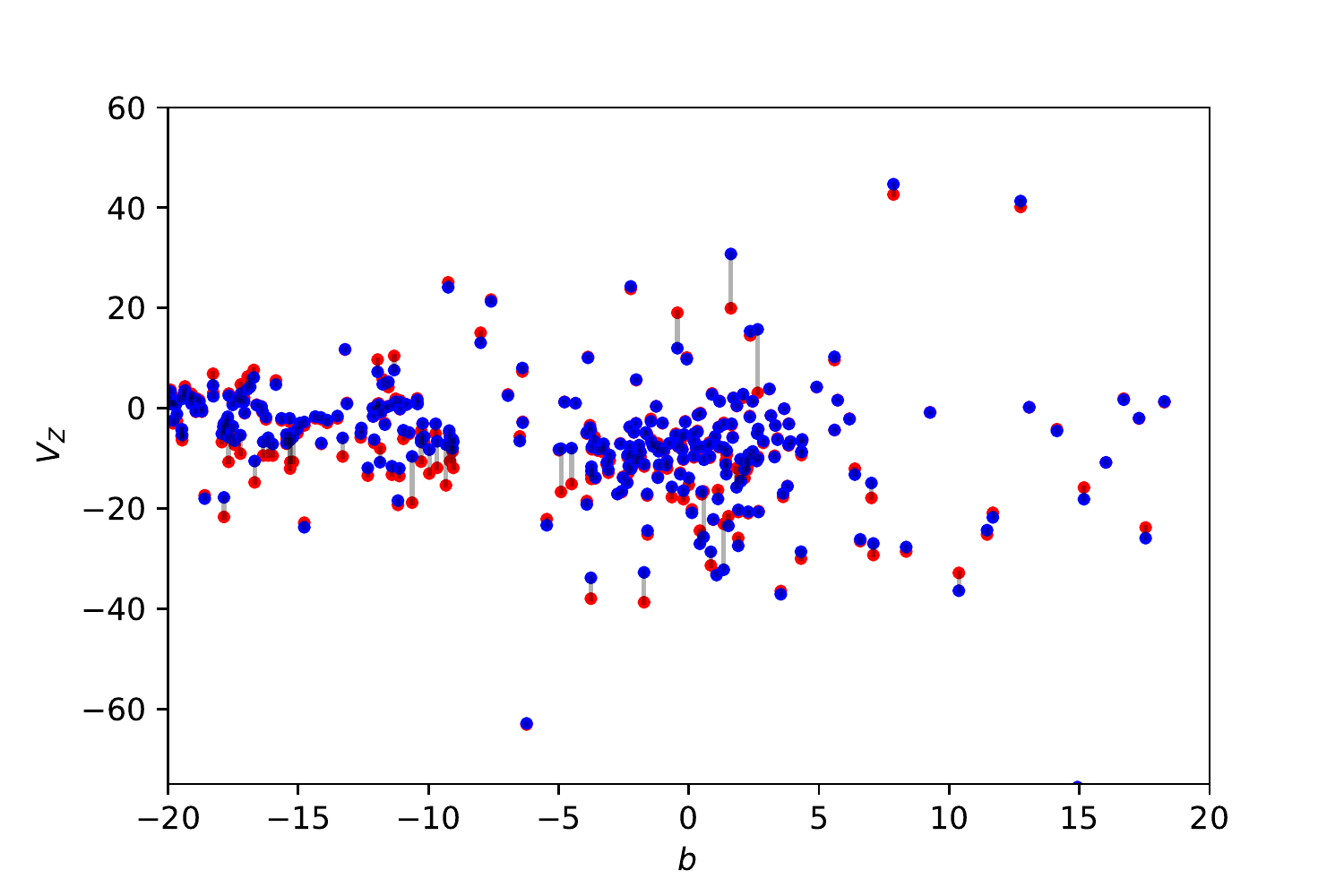}
  \caption { \label{fig:vzcomp} Vertical velocities $v_z$ of YSOs plotted against their Galactic latitudes $b$. The red and blue dots mark the vertical velocities that are derived with and without radial velocities, respectively, where the velocities of the same YSOs obtained from different methods are connected by grey lines. Only YSOs with LAMOST radial velocity measurements are plotted.}
\end{figure}

We compute the vertical velocity $v_z$ using the
\texttt{GalPy} package \citep{2015ApJS..216...29B}. To compute the 3D velocity, one would need the proper motions as well as the radial
velocities. In our case, radial velocity measurements {  are missing for} the majority of our YSOs.
However, {  this is not crucial, as we are only interested in $v_z$, and the} contribution of $v_r$ to $v_z$ scales with the ${\rm sin}(b)$ where $b$ is the Galactic latitude. If the source stay close the disk mid-plane, one can compute $v_z$ approximately by assuming $v_r = 0 $.
To verify our approach, we have matched the YSO candidates in our catalogue to those with
spectroscopic data from the LAMOST Data Release 7 (LAMOST
DR7\footnote{http://dr7.lamost.org/}; \citealt{Luo_2015}). In Fig.~ \ref{fig:vzcomp} we show a
comparison between the vertical velocities of stars derived with and without
radial velocities, where we estimate that the velocities of most of the individual YSOs differ by less than  20\,per\,cent. 

\end{document}